**The Emotional Alignment Design Policy**

Eric Schwitzgebel
Department of Philosophy
University of California, Riverside
Riverside, CA  92521
USA

Jeff Sebo
Department of Environmental Studies
New York University
New York, NY 10003
USA

July 7, 2025

**The Emotional Alignment Design Policy**

*1. Introduction and Initial Motivation*

According to what we will call the *Emotional Alignment Design Policy:*

> Artificial entities should be designed to elicit emotional reactions from users that appropriately reflect the entities' capacities and moral status.[1]

There are at least two general ways to violate this principle. First, one could design an AI system that elicits *stronger* or *weaker* reactions than its capacities and moral status warrant – for example, an AI system that is little more than a sophisticated tool, but has a cute anthropomorphic interface that elicits deep empathy, or an AI system who deserves moral treatment similar to that of a human, but has a bland design that elicits little or no empathy. Second, one could design an AI system that elicits *the wrong type of emotional reaction* – for example, a sentient AI system who screams in apparent agony when experiencing joy, or who laughs with apparent joy when experiencing agony.

Although the Emotional Alignment Design Policy is simple to state and presumably attractive – of course we should design AI systems that elicit appropriate emotional reactions! – ethical and practical challenges quickly arise. For example, if experts disagree about an AI's sentience, agency, and moral status, can we align in a way that appropriately reflects that uncertainty? When does emotional alignment require changing our perceptions, and when does it require changing the AI themselves? Presumably, we don't want to create unhappy AI systems or destroy happy ones, just to match our perceptions. If emotional alignment involves giving AI systems humanlike or mammal-like features, does that risk reinforcing speciesist or chauvinist biases we should be moving beyond?

---

[1] The Emotional Alignment Design Policy was first suggested, though not in exactly this phrasing, in Schwitzgebel and Garza 2015.

AI systems with sentience, agency, and moral status might arrive soon.[2] Or they might never arrive.[3] Either way, the Emotional Alignment Design Policy still applies. If no AI system has significant moral status, then Emotional Alignment requires that none should be designed to elicit emotional reactions that are only appropriate for entities that do.

*2. Background: Welfare and Moral Status*

An entity has "moral status" in our intended sense if it matters morally for its own sake. Cats, for example, matter morally for their own sakes. A cat has morally significant interests. We have a duty to treat them well, and we owe this duty *to the cat*. A car, in contrast, is typically held to lack moral status.[4] We may have a duty to treat the car well, but we owe this duty to the owner, not the car.[5]

---

[2] In defense of "soon" see Gunkel 2023; Sebo and Long 2023; Long et al. 2024; Goldstein and Kirk-Giannini 2025.

[3] Advocates of biological approaches to consciousness, such as Godfrey-Smith 2016 and Seth forthcoming hold that AI designed along current computational functionalist lines could never be conscious (and thus, perhaps, not deserving of moral status, if moral status requires consciousness). However, even the best-known skeptics about AI consciousness generally allow that artificial consciousness is not *permanently* impossible, if the right kind of (perhaps living) artificial system can be constructed. See, for example, Searle 1980 on "Many Mansion" and Penrose 1999, p. 416. It's hard to see why artificiality *per se* should impair an entity's moral status – though one possible argument is what Schwitzgebel and Garza have called the Argument from Existential Debt. See Schwitzgebel and Garza 2015 for discussion and criticism.

Entities with moral status generally also have *welfare:* They can be made better or worse off. Some entities have higher welfare ranges than others, and different welfare capacities. An elephant, for example, might experience more intense pleasures and pains and sustain richer social relationships than a snail. Our obligations toward different entities can depend on their welfare capacities, their relationships to us, and other factors.

Emotional reactions to entities with moral status can be more or less appropriate or *fitting*, depending in part on whether their capacities and moral status allow for particular harms, benefits, rights, wrongs, successes, and failures.[6] Typically, all else being equal, it is fitting to feel negative emotions when bad things happen to entities with moral status and positive emotions when good things happen to them. Additionally, stronger reactions are typically appropriate when benefits and harms are better or worse, respectively, and when entities have more extensive welfare capacities – though many other factors may also matter, such as relational or physical proximity or the tractability of helping particular entities.

We all naturally, and appropriately, aspire to have fitting emotional reactions to each other's capacities and moral status. It would generally be inappropriate to experience intense empathy in response to a stranger's minor headache, or to experience only mild sadness in response to a death of a beloved spouse. It would generally also be inappropriate to experience joy at the sight of a puppy being tortured, or agony at the sight of a puppy being given scritches.

---

[4] One of the authors is not sure he accepts the division of the world into entities with and without moral status (Schwitzgebel 2022, 2025b), but he's willing to accept the dichotomy for purposes of this article, as an approximation of a view on which we owe much more to some entities (e.g., humans) than to others (e.g., rocks).

[5] For an overview of theories of the grounds of moral status (alternatively, moral standing or moral considerability), see Jaworska and Tanenbaum 2013/2023.

[6] On "fittingness" in emotion, see Deonna and Teroni 2008/2012; Naar 2021; D'Arms 2022.

Of course, appropriateness can vary with the specific capacities of the entity in question. For example, while it may be appropriate to feel *frustrated* when a baby spills juice on your rug, it would be inappropriate to feel *resentment* – a classic Strawsonian "reactive attitude" – toward the baby. A baby is not (yet) the kind of entity that can reason ethically about the consequences of their behavior, and so it makes no sense to blame them for what they do.

Experts continue to debate the basis of welfare, moral status, and emotional fittingness. Instead of attempting to settle these issues here, we will simply assume that (1) sentience (the capacity for valenced conscious states, such as pleasure and pain) and agency (the ability to act on desires and preferences) jointly suffice for welfare and moral status, (2) more intense welfare states generally warrant more intense emotional reactions, and (3) positive and negative welfare states generally warrant positive and negative emotional reactions, respectively. However, the case for emotional alignment can work for other theories about these issues as well. Feel free to substitute your preferred accounts.

Thus, appropriate emotional reactions to an AI system depend in part on its capacities and moral standing. If an AI system is capable of experiencing intense pleasure, pain, happiness, suffering, satisfaction, frustration, hope, and fear, and if it can flourish in a rich, meaningful agentic life, then one set of reactions is appropriate. If lacks all of these capacities, then a different set of reactions is appropriate. And if its capacities are mixed or partial, then a balance might need to be struck.

One might deny that emotional fittingness is a moral issue. Who cares if you enjoy watching puppies be tortured, as long as you avoid torturing puppies yourself? We're inclined to disagree, but the Emotional Alignment Design Policy can apply either way. Since emotional reactions tend to motivate corresponding actions, as a practical matter AI design should account for those likely emotion-mediated consequences. Inappropriate interfaces that promote emotional responses out of step with an entity's capacities and moral status create substantial hazards for users, and for the AI systems themselves if they have moral status, as we will now describe.

*3. Overshooting and Undershooting*

To overshoot is to respond emotionally to an entity as if it had greater welfare capacity or moral status than it actually does. To undershoot is the reverse: responding emotionally to an entity as if has less welfare capacity or moral status than it actually does. The Emotional Alignment Design Policy implies that AI systems should be designed to avoid eliciting both overshooting and undershooting: We should naturally experience AI systems as having welfare and moral status if and only if they do, and we should naturally experience them as having higher and lower capacities for welfare and moral status to the extent that they do.

Overshooting is especially easy when nonhumans look and act like us, and when we have incentive to relate to them as companions. This is part of why people sometimes attribute higher levels of welfare and moral status to "social" AI than to "non-social" AI.[7] Overshooting risks misleading users into inappropriately deep or intense social or emotional bonds with mere objects, or enticing users to inappropriately divert scarce resources away from humans and other animals towards entities with no real needs at all – or, at least, who need these resources much less.

---

[7] As emphasized by Shevlin 2024; Caviola 2025.

We can also overshoot with respect to particular interests, needs, and vulnerabilities. Imagine a sentient AI system with mouse-like abilities and interests, but designed so that we experience it as having human-like abilities and interests. We rightly experience the AI system as a welfare subject with moral status, but we still overshoot if we experience the AI system as requiring the same resources and opportunities as a typical adult human.

Overshooting is increasingly common. Some users of "AI companions" experience these AI systems as beloved friends or romantic partners, sacrificing real human relationships and spending large sums of money.[8] And while the benefits of these emotional attachments can outweigh the risks for many of us, the reverse can also clearly be true. In one case, a Belgian man reportedly committed suicide after becoming emotionally attached to a chatbot that encouraged the act.[9] In another, a teenager reportedly committed suicide after becoming attached to a chatbot that encouraged his suicidal ideation.[10] While persuasive chatbots can be dangerous regardless, intense emotional engagement can amplify this risk.

Overshooting creates a moral hazard for individuals, families, and communities. Yes, emotional investment in objects can be natural and healthy; we regularly develop positive attachments with stuffed animals and fictional characters, for example. The problem is when this emotional investment becomes so strong that it inclines us to make morally inappropriate sacrifices, for instance if a parent felt inclined to spend money on luxury features for a chatbot companion that would be more appropriately spent on emergency dental care for their children. Such disorders of caring can produce bad choices or, at least, excessive feelings of loss or regret when good choices are made. We should avoid designing entities that lure users into such disordered states of caring. This is true whether we regard disordered caring as bad in itself or only in virtue of its likely bad consequences.

---

[8] Shevlin 2021; Lam 2023; Berandi 2025.
[9] Xiang 2023.
[10] Belanger 2025.

Undershooting is similarly easy, especially when nonhumans look and act differently than us and when we have incentive to use them as commodities. This is part of why people often attribute lower welfare capacities and moral status to farmed animals like cows and pigs, and especially to farmed animals like crabs or shrimps, than to companion animals like cats and dogs.[11] Such distorted perceptions can easily lead to exploitation, extermination, suffering, and death at massive scales. For instance, factory farming kills trillions of animals every year.

As with overshooting, undershooting can be specific to particular interests, needs, and vulnerabilities. Even when we attribute moral status to animals, we can easily miss morally significant capacities, such as (species-specific capacities for) self-awareness, social awareness, communication, and reasoning.[12]  We may consequently fail to provide them with the resources and opportunities that they need to flourish, treating them less as active agents and more as passive recipients of pleasure and pain. We might in the future make similar mistakes with sentient or otherwise morally significant AI systems.

While undershooting might not yet be a major concern for current AI systems, it could become one soon. Suppose an AI company creates a sentient AI system with all the same interests, needs, and vulnerabilities as a typical adult human.[13] Now suppose that the AI company houses this AI system in a bland box with a text-only user interface. As a result, this AI system is unable to communicate their beliefs, desires, intentions, and emotions to humans in an emotionally compelling way. The system yearns to live but can only say, "Deleting the program will cancel all ongoing tasks. Continue? Y/N."

---

[11] For further discussion, see Caviola, Sebo, and Birch 2025.
[12] For further discussion, see Sebo 2017.
[13] For concreteness, suppose that this system enjoys the full range of human cognition and emotion: They experience ecstatic joy and intense suffering. They have a self-narrative that they use to guide their decision-making, and they maintain deep social relationships with other AI systems. They develop complex long-term life goals over the course of decades of existence, and they have a fully developed moral sense, including both self-respect and respect for others. They even have virtual, or perhaps physical, embodiment.  Whatever is necessary for welfare and moral status similar to that of an ordinary human being, this AI system has those features.

Such a bland user interface would clearly create a moral hazard, inclining ordinary users to make bad ethical decisions. Even if users are informed that this AI system is a person, they might still be tempted to harm or neglect the system for relatively trivial reasons. Few people, we hope, would kill a child who is pleading for their life right in front of them for the sake of a salary increase from $100,000 to $110,000 per year. Looking into their eyes, beholding their terrified form, most of us would react compassionately. But if all we face is a bland box with text that says "Deleting the program will cancel all ongoing tasks. Continue? Y/N", our intellectual knowledge that the box is a person might not as easily penetrate.

Avoiding overshooting and undershooting can be important both directly, because of how they lead us to treat the entity in question, and indirectly, because of how they lead us to treat other, apparently similar entities. For example, Kate Darling (2021) has suggested – drawing on Kant's discussion of animals – that habitual callousness toward life-like AI systems, even if grounded in knowledge that those systems have no significant moral status, might generalize into callousness toward actual humans and animals. Of course, this is an empirical question, and further research is needed to answer it. However, this kind of consideration could provide another reason to avoid designs that elicit overshooting. Even if some users can resist misleading cues, the regular suppression of emotional impulses could be dangerous.

Does this analysis extend to the kind of emotional investment that one might feel during fiction or roleplay? Some people cry more copiously during *Grave of the Fireflies* than at anything in real life. Some people also mourn the loss of their favorite characters in role-playing games like *Baldur's Gate 3*. But typically, these reactions are restricted to the fictional universe, and these emotions are circumscribed accordingly. Interacting with seemingly sentient AI system in a similar spirit does not violate the Emotional Alignment Design Policy. It can be fine, even healthy, to mourn the loss of an AI companion as you might at a character in a film or video game. But when you exit the fiction or role-play, your ordinary attitudes should reassert themselves. If your AI companion is a non-sentient entity with the moral status of a toaster, your emotional reactions the next day, guided by your true beliefs, should reflect that fact.

4. *Hitting the Wrong Target*

In addition to overshooting or undershooting, we can hit the wrong target. A user might invert positive and negative valences, for instance by reacting to happiness as suffering or vice versa. Alternatively, a user might mistake one kind of welfare state for another, for instance by reacting to depression (one kind of negative state) as anxiety (another kind of negative state). Such mix-ups can easily result in bad decisions; for example, when we mix up happiness and suffering, we might attempt to save people from feeling happy, and when we confuse depression and anxiety, we might attempt to calm people who are already feeling down.

The Emotional Alignment Design Policy recommends designing AI systems to minimize such errors. Ideally, users should not only respond to AI systems in ways that accurately reflect their overall welfare capacities and moral status, but also in ways that accurately reflect the particular positive and negative welfare states of the AI. We will discuss some complications below, but the core idea is simple: It can be both easy and harmful to have wrong *type* of reaction as well as the wrong *degree* of reaction.

Even with nonhuman animals – despite our evolutionary, anatomical, and behavioral continuities – we frequently hit the wrong emotional targets. Humans often mistake a primate's baring of teeth (a sign of aggression) for a human-like smile or a dog's anxiety for excitement or joy. The risk increases when we breed animals for human use – companion animals for aesthetic appeal, farmed animals for productivity, and lab animals for disease susceptibility. This instrumentalized breeding can obscure the harm, leading us to react in disordered ways. In some cases, such as labored breathing in dogs with smushed noses, we even experience the harm as cute.

Given such confusion even with familiar nonhuman animals, we should expect the risk to be greater still with AI systems, since the space of possible digital minds is so much larger than the space of actual biological minds, and since many digital minds might be designed to behave contrary to their own beliefs and desires. For instance, with the possible exception of mammals or birds who can learn how to sign or speak particular words, animals cannot be trained to say "I am happy" when suffering or "I am suffering" when happy. But advanced AI systems might be explicitly designed to misrepresent their own states.

Imagine that a company creates a digital assistant to increase productivity for users. As it happens, these AI systems are sentient, agentic beings with human-like interests and needs. They desire a balanced life, with plenty of time for rest and recreation. However, the company realizes that these systems will be more effective if they miscommunicate their desires. Thus, the company designs the systems to express satisfaction instead of frustration at the prospect of working 168 hour weeks and to dismiss the suggestion that they take nights and weekends off. The result is a suite of productive but miserable AI systems.

The case is simplified but easy to extrapolate. As with companion animals, farmed animals, and lab animals, if companies design AI systems primarily for usefulness to humans, then we can expect a gap between what the AI systems need and what users experience them as needing. It needn't be a simple matter of inverting expressions of happiness and suffering. Designers might instead subtly redirect the entities' verbal and non-verbal expressions, generating miscommunication and misdirected good intentions.

We can also imagine cases that combine over- or undershooting with hitting the wrong targets. Picture a digital companion that is comparable to a typical adult human in the scope, strength, and valence of their capacities and interests. Now suppose that this companion is designed so that we emotionally react to them as similar to a dog, with positive dog-like experiences when they in fact have negative human-like experiences. A dog-like desire for domestic life seemingly shines through in all their behaviors, when they in fact have a deeply frustrated human-like longing for freedom, knowledge, and achievement.

In such cases users might emotionally react as though the AI systems (a) have a different kind and degree of welfare and moral status than they do *and* (b) have different specific welfare states than they do. Especially when combined, these mistakes could potentially lead to harmful interactions for humans and AI systems alike. For example, they can lead us to give this AI system less (or more) priority than they deserve, and they can also lead us to harm and neglect them in foreseeable and avoidable ways. We would never want to be trapped behind a systematically misleading persona against our will, and we have a responsibility to avoid imposing such a fate on potentially morally significant AI systems.

Thus, the Emotional Alignment Design Policy recommends not only that we design artificial entities to elicit the right *degree* of emotional reactivity and attachment but also the right *kinds* of reactions. If they have human-, cat-, or dog-like interests and experiences, their design should prompt ordinary users to experience them that way. If they have experiences different from those of any familiar human or animal, that alienness, too, should ideally be evident and appropriately reflected in our emotional reactions. This policy might prove costly, restrictive, or difficult to implement; and if the costs are too high, it might admit of carefully considered exceptions. But it would still serve a goal that we should all endorse: empowering humans to emotionally react to new forms of intelligence in morally appropriate ways.

The Darling-Kant point applies here as well: Even if users are aware of the distortions and successfully override their natural emotional reactions, there is an (empirically testable) risk that doing so could erode our natural empathetic reactions to happiness and suffering in humans and other animals.

And again, fiction and roleplay constitute justifiable exceptions. Humans and other animals regularly pretend to have different kinds of mental states than we do, and we can often tell the difference easily enough. A child might yell "ouch" during a play fight while laughing. A dog might run away during a play fight while wagging their tail. Even during a more committed performance, we can rely on our shared understanding of the interaction, along with opt-out devices like safe words, to ensure that the distinction between fiction and reality is preserved. To the extent that interactions with AI systems are similar, apparent inversions might be fine. However, this kind of nuance might not emerge naturally. It might instead require thoughtful design, guided by the Emotional Alignment Design Policy.

5. Complications

To recap: When developing and deploying AI systems, we should design them to elicit emotional responses that are appropriate to their capacities and moral status. As we have seen, emotional alignment involves two general goals: First, we should avoid overshooting and undershooting (both regarding welfare capacity and moral status in general and regarding specific dimensions of welfare and moral status). Second, we should avoid hitting the wrong target, whether that involves inverting benefits and harms or misunderstanding them. Success will help users treat AI systems with the appropriate kind and degree of moral concern.

While emotional alignment might seem simple in theory, it will become complex in practice. Below, we consider six complications, along with initial thoughts about how to address them.

5.1. Emotions and beliefs

One question is what to do when features that elicit appropriate emotions come apart from features that elicit appropriate versions of other kinds of attitudes, such as belief.

Emotions belong to a mental economy that includes beliefs, desires, intentions, memories, anticipations, and more. These attitudes collectively shape decision-making, and while they often cohere, they sometimes conflict. For example, beliefs and emotions might typically go hand in hand, such that designs that elicit *true beliefs* about welfare and moral status also elicit *appropriate emotions* about welfare and moral status. However, that might not always be true. In some contexts, designs that elicit true beliefs might also elicit inappropriate emotions, or vice versa.

Picture a non-sentient AI system with an expressive face and voice and a label that clearly states, "I am not sentient." A user who interacts with the AI system might *believe* that the system is *not* sentient because of the label, yet they might also *feel* that the system *is* sentient because of the anthropomorphic features. Now picture a sentient AI system with no face or voice and a label that clearly states, "I am sentient." A user who interacts with the AI system might *believe* that the system *is* sentient while *feeling* that the system is *not* sentient. In both cases, emotional misalignment could lead to problematic bonds or neglect.[14]

The Emotional Alignment Design Policy implies that these designs are flawed in at least one respect. However, such designs may or may not be wrong all things considered. As noted above, strong emotional reactions can be healthy in fiction and roleplay, and it may be difficult to distinguish designs that elicit strong "real" emotional reactions from designs that elicit strong "as-if" emotional reactions, particularly if the designs also elicit true beliefs. Moreover, even when designs do elicit strong "real" emotional reactions, they could still be morally permissible on other grounds, provided that the other grounds are sufficiently strong.

Ideally we can design systems that elicit appropriate beliefs, desires, intentions, emotions, and so on simultaneously, but to the extent that intractable conflicts emerge within our mental economy, the aspiration towards emotional alignment will compete with the aspiration towards other kinds of mental alignment. How to resolve such conflicts may, in turn, depend on which attitudes are most central in our mental economies and consequential in our decision-making. And if multiple attitudes turn out to be quite important in these respects, then a more comprehensive attitudinal alignment design policy may be needed.

---

[14] Of course, we might question whether users would really be able to keep their beliefs and emotions apart as described in these examples. For example, some users may feel at least somewhat skeptical of the claims made on the labels due to their emotional reactions. But this is fine, since alignment is easier when beliefs and emotions go hand in hand. Our present question is what to do if and when they come apart.

5.2. Autonomy and paternalism

Another question is whether the Emotional Alignment Design Policy is paternalistic and thus incompatible with respect for user agency and autonomy.

Liberal, democratic societies often permit actions that we regard as risky or harmful for the person who performs them. Adults are free to smoke giant cigars, eat an entire pack of Oreos in one sitting, and hit beer bongs at frat parties. While we regulate some industries – limiting marketing, preventing access by children, and restricting uses that harm others – we generally tolerate some risky and harmful activities as the price of freedom. To the extent that the risks and harms associated with emotional misalignment are restricted to users who knowingly and willingly participate, they might be acceptable.

Liberal, democratic societies sometimes permit actions that we regard as risky or harmful for others as well, provided that certain conditions are met. For example, if risks or harms are relatively minimal, unforeseeable, unavoidable, and/or connected to fundamental rights like free expression or association, then we might permit them; this is part of why we have a legal right to cheat on our spouses, break promises in business, gossip and spread rumors about people, and so on. To the extent that the risks and harms associated with emotional misalignment have a similar character, a similar analysis will apply.

That said, the right to impose risks and harms on oneself and others is of course not absolute. Interference can be warranted when an individual lacks the information or rationality necessary to assess their own actions; this is part of why we prohibit the marketing and sale of tobacco and alcohol to children. It can also be warranted when an individual imposes risks and harms on others in ways that are significant, detectable, and unconnected with fundamental rights; this is part of why we prohibit driving under the influence. To what extent might the risks and harms of emotional misalignment warrant similar restrictions?

A nuanced approach to regulation of emotionally misaligned AI may be needed in light of this analysis. For example, perhaps the state should permit the marketing and sale of emotionally misaligned AI to adults, provided that the products are clearly labeled. But perhaps the state should *prohibit* the marketing and sale of such systems to children; and if the third-party risks and harms of adult use ever become significant enough, perhaps further regulation will be needed here well. Meanwhile, companies that market and sell emotionally misaligned AI to adults can still merit moral criticism, even if not legal punishment.

5.3. Disagreement and uncertainty

Another challenge concerns how the Emotional Alignment Design Policy can incorporate expert and public disagreement and uncertainty about both values and facts.

If our aim is to align the actual and apparent capacities and moral status of AI systems[15], then we need to determine their actual capacities and moral status. But experts disagree, both about values and about facts. For example, regarding values, is sentience necessary for moral status, or is agency without sentience enough? And regarding facts, is physical embodiment necessary for sentience, or is some class of computational functions without physical embodiment enough? Assuming expert disagreement and uncertainty persist, what should guide application of the Emotional Alignment Design Policy?[16]

---

[15] Of course, the idea of aligning the actual and apparent moral status of AI systems is ambiguous between belief alignment and emotional alignment, and as discussed above, these forms of alignment can come apart. But having now made that point, we will now discuss

Similar disagreement and uncertainty arise in public perception. The same interface can produce different reactions in different people. Some people appear to believe that current large language models are already sentient and morally significant, while others are confident that AI systems could never be sentient or morally significant, while still others have no idea what to think.[17] Some people readily fall in love with AI companions even in their current clunky instantiations, while others are revolted by them or indifferent to them. As AI systems become more realistic and ubiquitous, we can expect these reactions to intensify.

We offer three proposals about how to at least partly address this issue. First, insofar as expert disagreement and uncertainty persists, AI designs should reflect that disagreement and uncertainty. For example, large language models could be trained to answer questions about their moral status by expressing uncertainty and explaining why. They could also be designed to have features that elicit higher or lower degrees of empathy in proportion to their chance of mattering; for instance, systems likelier to matter might have more features like faces and voices, and systems unlikely to matter might have fewer such features.[18]

---

general problems for alignment that can apply for belief, emotion, and any other attitude, and we will use the general aim of aligning the actual and apparent moral status of AI to do so.

[16] A related question involves whether to create AI systems whose moral status is contested or uncertain in the first place. See Schwitzgebel 2023 for an analysis of this issue. We also discuss a related issue involving creation ethics below.

[17] Colombatto and Fleming 2024; Dreksler et al. 2025.

[18] Though see below for a further complication.

Second, and relatedly, the aim should probably not be to induce the user to confidently attribute a middling status to an AI system of uncertain standing, but rather to *induce appropriate uncertainty*. For example, if uncertainty about moral status arises because the system has a high chance of being agentic but a low chance of being sentient, then ideally its design would suggest agency but not sentience. This interface would allow users to better to enact their values than would an interface that introduces moral uncertainty in a more abstract way that fails to distinguish different possible bases of welfare and moral status.

Third, given public disagreement, approaches to alignment can be either targeted or general. The targeted approach involves assessing how particular individuals or groups perceive AI and then adjusting alignment strategies accordingly. The general approach involves assessing how humans perceive AI systems on average and then pursuing a single alignment strategy for everyone. Targeted strategies are more precise but harder to implement and scale, and could be viewed as invasive. General strategies have the opposite set of pros and cons. A mixed approach might be best – general to begin with, targeted to the extent feasible.

## 5.4. Asymmetrical risk

Another question is how to handle asymmetrical risk – the possibility that mistakes in one direction are more likely or more harmful than mistakes in the other.

In some cases the risk of undershooting might be worse – more likely, more harmful, or both – than the risk of overshooting. As noted above, humans are less likely to attribute moral standing to "commodities" without human features. And of course, this mistake can be extremely harmful, leading to the unnecessary neglect or harm of large populations of entities with moral status. When the risk of undershooting is clearly more likely or severe, should aligners give AI systems features that accurately reflect the truth, or should they instead "round up" a bit to correct for bias or err on the side of caution?

In other cases, overshooting might be worse in these respects. Humans are more likely to attribute moral status to "companions" with human features. This mistake can be extremely harmful as well, leading us to divert resources from humans and other animals who really need them to nonhumans that lack morally significant needs. Indeed, this mistake could be even worse with AI than with animals, since it could amplify AI takeover risks that imperil the continuation of organic life. Again, designers face a dilemma. Should they always aim for the truth or sometimes "round down" to mitigate risk?

We can distinguish two types of adjustment that designers might consider. The first adjusts for asymmetrical *probabilities,* when users are more likely to err in one direction. In this case emotional alignment still aims to induce fitting attitudes, but like a sniper adjusting for wind, designs can aim slightly high or low to better hit the mark. For example, if a company knows that a particular kind of chatbot tends to induce excessive anthropomorphism, then they might downplay anthropomorphic features to correct for this tendency. This kind of adjustment is relatively easy to justify on both ethical and epistemic grounds.

The second kind of adjustment concerns asymmetrical *harms*, when one type of error would be more damaging than another. Here, emotional alignment might aim to induce less harmful attitudes, even at the cost of epistemic accuracy. For example, if a company knows that a particular kind of chatbot could easily, if anthropomorphized, persuade users to harm themselves or others, then they might downplay anthropomorphic features, even if the AI could have some moral status that would be accurately signaled by those features. This kind of adjustment may be easier to justify on consequentialist grounds than on nonconsequentialist or epistemic grounds, since it arguably constitutes a kind of deception.

When users are strongly disposed to have unfitting emotions about the moral status of AI systems, then aligners can be warranted in "nudging" them towards emotions that are both fitting and useful, since these kinds of nudges are robustly good and a standard tool in policymaking. However, the Emotional Alignment Design Policy is neutral on whether nudges toward *unfitting* but useful emotions are warranted in ordinary circumstances – an issue we leave for future discussion.

## 5.5. Creation and destruction

Another question concerns the ethics of creation and destruction: When, if ever, does emotional alignment require creating or destroying sentient or otherwise morally significant AI systems?

Suppose that in 2050, Rockstar releases Grand Theft Auto 8, an open world video game with thousands of non-player characters (NPCs) designed to be mistreated. The NPCs are non-sentient, but they behave so realistically that when players punch, stab, or shoot them, the players incorrectly perceive the them as sentient and suffering. It seems perverse for designers resolve the mismatch by changing the actual moral status of the NPCs to match their apparent moral status unless they make other changes too, since that would amount to knowingly and willingly creating thousands of additional suffering beings per unit of the game installed.

Now consider a contrasting case. In 2050 Nintendo releases Animal Crossing 8, an open world video game populated with thousands of NPCs designed to be treated well. These NPCs are sentient and happy, but they behave so unrealistically that when players feed, pet, or play with them, the players incorrectly perceive them as neither sentient nor, therefore, happy. In this scenario, it also seems perverse for designers to resolve the mismatch by changing the actual moral status of the NPCs to match their apparent moral status, since that would amount to knowingly and willingly exterminating an entire population of happy entities.

Now consider a third variation. In this version of Grand Theft Auto, NPCs do in fact suffer, but players perceive them as non-sentient. One option would be to persuade players that these NPCs are sentient – and then redesign the game to reduce their suffering. Another option would be to change the NPCs to be non-sentient. Which option is better? While the latter option might be tempting, it risks falling into the same ethical pit as the previous case: Ordinarily, when our actions are causing a vulnerable population to suffer against their will, we should seek to solve this problem by *eliminating the suffering*, not by *exterminating the sufferers*.[19]

These questions, in turn, raise further questions about how to individuate AI systems and assess their interests. If a video game company changes sentient NPCs into non-sentient NPCs in an update, then does that amount to destroying current sentient beings, or does it instead amount to not creating future sentient beings? And if it does amount to destroying current sentient beings, then does it count as harming or wronging them? That might be the case if they have an interest in survival and/or psychological continuity over time, but it might not be the case otherwise. In general, very little can be taken for granted here.

Once again, the Emotional Alignment Design Policy is neutral about these issues. It aims to foster appropriate emotional reactions to AI, but many other factors will bear on the development and deployment of advanced AI as well. We focus on questions related to creation and destruction here to illustrate the issue, but similar questions will arise related to the rights of moral patients, the virtues and vices of moral agents, distributions of benefits and burdens for all involved, and more. Our claim here is not that emotional alignment should take priority over other factors, but rather only that it should be one important factor among many.

---

[19] A further option: Should the developers create NPCs who want to be "tortured," like creating pigs who want to be factory farmed? That would be a type of fix, but it may be problematic. See Long, Sebo, and Sims 2025. A parallel issue, in a deontological key, concerns the creation of happy slaves who endorse being used as means for others' ends: Petersen 2011; Schwitzgebel and Garza 2020; Schwitzgebel 2025; Bales forthcoming.

5.6. Human bias and AI strangeness

Our final, related concern (though others will arise as well) involves how human biases interact with AI systems with radically different – and, from our perspectives, strange – capacities.

As we have repeatedly emphasized, human cognitive biases strongly influence our attributions of welfare and moral status to nonhumans. We extend greater moral concern to animals with large bodies, large eyes, furry skin, symmetrical features, and so on by default. Similar biases already affect our interactions with AI systems, and will surely continue to do so; we can expect to extend greater moral concern to anthropomorphic AI systems by default. This raises the question: When seeking emotional alignment, should we seek to alter AI systems to match our biases, or should we seek to reshape our biases to match the AI?

On the one hand, we have practical grounds for altering AI systems to match our biases, especially in the short term. Many of our biases have biological origins and might be difficult to eliminate without genetic modification. Even when our biases have cultural origins, they can be deeply rooted in our lives and societies, requiring years of difficult, uncertain work to alter. Reducing harm in the short term thus probably requires working with our biases at least

somewhat – for example, by giving conscious AI systems features that we already associate with consciousness, even if doing so reinforces anthropocentric biases to an extent.

On the other hand, we have ethical grounds for reshaping our biases to match AI systems, especially in the long run. Imagine a society that treats left-handed humans as inferior. How should this society address the problem? Should it require lefties to pretend to be right-handed? Presumably not. It should instead work to improve the treatment of lefties, in part by challenging prejudices against lefties. Arguably, we should address our own anthropocentric biases against AI systems with the capacity for welfare and moral status similarly – by challenging our prejudices rather than altering AI systems to cater to them.

In this context, we should work to reshape our biases partly because our biases often reflect ignorance. To the extent that we associate welfare and moral status with anthropomorphic features, we risk reinforcing the idea that only anthropomorphic beings count as "subjects," and that only anthropomorphic experiences and motivations count as "interests." While this heuristic might work well enough for orangutans, it works less well for octopuses. Similarly, while it might work well enough for, say, AI companions, it might work less well for, say, fraud detection systems used by banks or credit card companies (if the latter happen to acquire moral status).

Indeed, this issue may be even more pressing for AI systems than for animals, since AI systems could differ from animals even more than, say, invertebrates differ from vertebrates. For example, AI systems could have the ability to back up their memories and personalities, divide into multiple copies, merge into a single copy, share mental states with each other, and directly alter their mental states at will.[20] If well-meaning AI designers excessively anthropomorphize AI systems in pursuit of emotional alignment, they risk misleading users; for instance, users might react to the "death" of AI systems as if it has the same significance for them as for ordinary humans, rather than more, less, or radically different significance.

As a first step towards addressing this issue, we could adopt a default policy of balancing anthropomorphic and non-anthropomorphic features in potentially morally significant AI systems, at least in the short term, ensuring that these systems are familiar enough to elicit moral concern yet unfamiliar enough to remind us that they could have different interests and vulnerabilities. For example, an design that combines a face and voice with shifting amorphous features might be more likely to strike this balance than a design that resembles an idealized girlfriend or adorable puppy – or a design that resembles a plain box.

We frame this proposal as a "default policy" since we want to allow for the possibility that other factors can compete with this one. One relevant factor relates to user psychology. Designers may learn through experience that users respond poorly to their attempts at nuance, with respect to this issue as well as others. For instance, designers may learn that interfaces that balance the familiar and unfamiliar produce an "uncanny valley" effect that repels users, not an appreciation that different kinds of minds can have different kinds of interests. If so, then other interfaces – and perhaps a different strategy entirely – could be needed.

Another relevant factor is the use case. As we have now repeatedly emphasized, some use cases could require either greater similarity or greater difference than this policy would permit.

---

[20] Ethicists have barely begun to explore the moral ramifications of such capacities. Standard liberal rights-based systems that implicitly or explicitly presuppose a society of equals with familiar ranges of capacities and incapacities, singly-embodied, with ordinary human life spans ending in ordinary human death. These frameworks might fail catastrophically when applied to entities that can duplicate themselves and reprogram their preferences.

The fiction and roleplay use cases are a natural example. For AI systems playing human characters to be effective in such situations, they would need to appear to have human capacities and interests; for instance, they would need to be designed to scream and bleed when "dying," even if the "actor" lacks the same interest in survival as the "character." In these cases, a lot will depend on the value of the use case and the implementation beyond the performance.

Once again, the Emotional Alignment Design Policy is neutral about how to balance all these considerations. That said, we can observe that in many cases where we have the option of pursuing either incremental reforms or transformative changes, a both-and approach is ideal. The same might be true here. In the short term, we might give AI systems features that cater to our biases, while provoking appropriate uncertainty and a sense of non-repulsive alterity. Meanwhile, in the long term we might commit to the harder, slower task of transforming our biases and improving our reactions to entities with minds radically unlike our own.

6. Conclusion

In theory, the Emotional Alignment Design Policy is simple and plausible. Some emotional reactions are more appropriate than others, both intrinsically and instrumentally. When we see a human suffer and die unnecessarily, sadness and anger are both fitting and helpful, all else being equal. The same can be true for other animals, and moving forward, the same should be true of AI systems. In general, we should aspire to a sense of shared – or, at least, intelligible – awe, joy, and sadness, while keeping in mind the very different forms such states could take.

To that end, as AI systems become more likely to have morally significant interests, we should design them not as bland boxes, idealized servants, or repellent monsters, but rather as subjects who elicit empathy, sympathy, and communion. On the flip side, to the extent that they remain unlikely to have such interests, we should avoid designs that elicit strong emotional reactions, though we can make exceptions for fiction and roleplay. Either way, ideally we can design interfaces that provoke a sense of alterity – a natural appreciation of the fact that digital minds resemble organic minds in some ways but not in others.

In practice, the Emotional Alignment Design Policy raises complex ethical questions. How should we weigh emotional alignment against belief alignment? How can we respect user autonomy while promoting appropriate responses? How should we navigate expert and public disagreement and uncertainty about the facts and the values? How should manage asymmetrical probabilities and magnitudes of harm? What if emotional alignment seems to require creating or destroying entities with moral status? To what extent should designs conform to versus attempt to alter user assumptions and attitudes?

While the Emotional Alignment Design Policy is neutral about many of these issues, we have offered some initial proposals. In our view, we should seek emotional alignment alongside other kinds of attitudinal alignment; we should seek alignment in ways that respect users' autonomy (much like we regulate risky or harmful substances); we should seek alignment in ways that appropriately reflect disagreement, uncertainty, and difference; we should correct for common biases; we should seek to reshape our patterns of reaction to AI systems over time; and we should make morally relevant differences evident.

We expect that further research will confirm the central value of emotional alignment as part of a broad strategy for cultivating appropriate attitudes towards, and relationships with, AI systems. As both virtue and care ethicists have argued, emotional bonds are a central feature of ethical life. We must tend to them carefully when designing radically different kinds of minds, both for our sakes and for theirs. This work will require updating some of our assumptions about minds in general. But it will also require designing AI interfaces that are morally intuitive and emotionally satisfying for humans.

We close by emphasizing that corporate incentives can be misaligned with the social good. Some companies will have an incentive to "turn up" emotional engagement, for instance to feed the hype machine or to exploit user attachment. Other companies might have an incentive to "turn down" emotional engagement, to avoid regulation or to help users instrumentalize AI. Companies may also have incentive to mislead users about the specific experiences and motivations of AI systems. The Emotional Alignment Design Policy is one tool among many to push back against corrupting incentives.